\documentclass[12pt]{article}
\usepackage{graphicx}
\usepackage[utf8]{inputenc}  
\usepackage[T1]{fontenc}
\usepackage{amsfonts,amssymb,amsthm,amsmath}
\usepackage[french,english]{babel}
\usepackage{graphics,graphicx}

\begin{document}
\title{An Analytical Expression for the Hubble diagram of supernovae and gamma-ray bursts}
\author{J.-M. Vigoureux, B. Vigoureux\\
Institut UTINAM, UMR CNRS 6213,\\Université de
Franche-Comté, 25030 Besançon Cedex, France.\\
{\em jean-marie.vigoureux@univ-fcomte.fr}\\\\
M. Langlois\\
Passavant, 25360, France\\ 
{\em mj.langlois@wanadoo.fr}
}
\maketitle %\fontsize{12}{24}\selectfont

\begin{abstract}
\noindent A recent paper by Harmut Traunmüller shows that the most adequate equation to interpret the observations on magnitude and redshift from 892 type 1a supernovae would be  $\mu = 5\,log[(1+z)\,ln(1+z)] + const.$ We discuss this result which is exacly the one we have obtained few years ago when postulating a relation between the speed of light and the expansion of the universe. We also compare our analytical result to the conclusion of Marosi who studied 280 supernovae and gamma-ray bursts
in the range 0.1014 < z < 8.1. The difference between his results and ours is at worst of $0.3\, \%$.
\end{abstract}
PACS numbers: 98.80.Cq, 97.60.Bw, 98.80.Es, 98.70.Rz, 98.62.Py \\
KEYWORDS : supernova type Ia-standard candles, gamma ray burst experiments, cosmology, redshift/magnitude data, universal constant, speed of light, cosmological model, cosmology: observations, cosmology:theory, supernovae.
\section{Introduction}
Some years ago we proposed \cite{Vigoureux08} \cite{Vigoureux03} \cite{Vigoureux} \cite{Viennot09} that the constant $"c"$ (the speed of light) and the expansion of the universe are two aspects of one single concept connecting space and time in the expanding universe by putting  (with coordinates normalized to the present epoch)
\begin{equation} \label{c}
c = \alpha \frac{dR(t)}{dt} =\alpha \,R_0\, \frac{da(t)}{dt}= Const.
\end{equation}
where $\alpha$ is a constant, where $a(t)$ is the cosmic scale factor and where it must be emphasized that $c$ is constant. This equation (\ref{c}) lead us to propose a cosmological model one of its interests being to drastically reduce the number of problems of standard cosmology.
Among our results, we showed that eq.(\ref{c}) leads to interpret the distance modulus vs. redshift curve \textit{without having to consider any acceleration of the universe}. Recently
quite independently and from statistical considerations, Harmut Traunmüller \cite{xxxx}, using data on magnitude and redshift from 892 type 1a supernovae showed that a statistical study suggests that for standard candles, "magnitude $m = 5\,log[(1+z)\,Ln(1+z)] + const.$ gives the best fit of all results" so that
\begin{equation} \label{HT}
\mu = \text{Const.}  +5 \log{ \left((z+1) \ln{(z+1)}\right)} 
\end{equation}
This last expression being exactly the one we obtain \cite{Vigoureux08} with (\ref{c}), we come back to this result.
We also show that our result agrees with a high precision with the one obtained  in the range of $z =0.0104$ to $8.1$ by Marosi \cite{Marosi14} who compared the Hubble diagram calculated from the observed redshifts data of 280 supernovae  with Hubble diagrams inferred on the basis of different cosmological models.
\section{An analytical expression for the Hubble diagrams}
To calculate the expression for the distance modulus $\mu$ with respect to $z$ let us consider an 
object at cosmic coordinate $\chi$ and let us suppose that the light that is emitted at cosmic time $t_e$ is just reaching us at time $t_0$. Let us also write the Robertson-Walker metric (with coordinates normalized to the present epoch and $R(t)=R(t_0)\, a(t)$) in the form
\begin{equation}
ds^2=- c^2 dt^2 + R(t)^2 \left(d\chi^2 + S_k^2 \,d\Omega^2 \right) 
\end{equation}
Using (\ref{c}) which obviously gives $R(t_0) = R_0= c \,\dfrac{t_0}{\alpha}$ and $H_0= \dfrac{\dot{a}(t_0)}{a(t_0)} = \dfrac{1}{t_0}$ ($H_0$ is the Hubble constant at time $t_0$) the luminosity distance $d_L$ of the object can be expressed as 
\begin{equation}\label{dL1}
d_L=(z+1)\,R_0 \, \chi= \frac{c}{H_0 \, \alpha}\,(z+1)\, \chi
 \end{equation}
$\chi$ can be obtained by writing that light travels on a radial null geodesic so that $c\, dt = R(t)\, d\chi$. Using (\ref{c}) we thus have:
 \begin{equation}
\chi =\int_{t_e}^{t_0}\frac{c\, dt}{R(t)} =\int_{t_e}^{t_0}\frac{\alpha\, da(t)}{a(t)}= \alpha \ln{\frac{a(t_0)}{a(t_e)}}
\end{equation}
Introducing this result into eq.(\ref{dL1}) with $\dfrac{a(t_0)}{a(t_e)}= z + 1$, gives
\begin{equation}\label{dL}
d_L=\frac{c}{H_0}(z+1) \ln{(1 + z)}
\end{equation}
The distance modulus is related to the luminosity distance via 
\begin{equation}
\mu = 5\, \log{\left(d_L(Mpc)\right)}+25
\end{equation}
so that introducing $d_L$ into that equation gives
\begin{equation}\label{mu}
\mu = 25 + 5\, \log{ \left(\frac{c}{H_0}\right)}+5 \log{ \left((z+1) \ln{(z+1)}\right)} 
 \end{equation}
That equation which we found in \cite{Vigoureux08} is exactly the one that Harmut Traunmüller \cite{xxxx} found quite independently as being the best to interpret the observations of 892 type 1a supernovae.\\
\begin{figure}[h]
\centering
\includegraphics[scale=0.6]{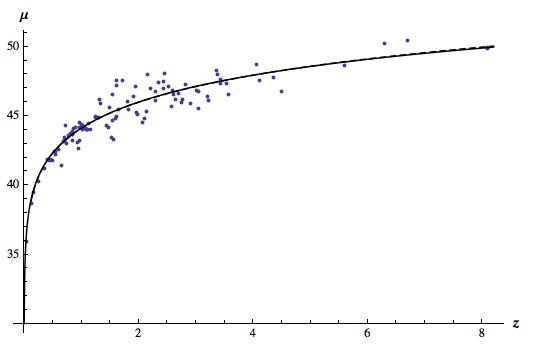}
\caption{Hubble diagram (distance modulus $\mu$ vs. redshift $z$). The unbroken line represents predictions from eq.(\ref{c}) and (\ref{mu}) (We take $H0= 62.5 \, km\,s^{-1}\,Mpc^{-1}$). All data points are taken from Hao Wei paper \cite{Wei10}. The dotted line (which is barely visible because it is so close to our result) corresponds to the best function ($\mu = 44.109769 \,z^{0.059883}$) obtained by Marosi \cite{Marosi14}. \label{fig 1}}
\end{figure}
\noindent Fig(1) shows our result for $\mu$ vs.\,$z$ (all data points are taken from the paper of Hao Wei \cite{Wei10}).
It exactly corresponds to the one (\ref{HT}) of  Harmut Traunmüller \cite{xxxx}. It also agrees with the one of Marosi \cite{Marosi14} (dotted line in fig.1)
who, comparing the Hubble diagram calculated from the observed redshift data of 280 supernovae with Hubble diagrams inferred on the basis of two cosmological models in the range of z = 0.0104 to 8.1, found that the best way to represent observations would be expressed by 
\begin{equation} \label{mub}
\mu = 44.109769 \,z^{0.059883}
\end{equation}
Calculating the difference between (\ref{mu}) and (\ref{mub}) in fact shows that, at worst, his numerical values agree with ours within at worst $0.4\,\%$ (this worst value being obtained for $z = 8.1$). This agreement explains that Marosi's result is barely visible in fig.1 (dotted line) because it is so close to our result. 
\section{Discussion}
Although the supernovae results are interpreted in the standard cosmology as indicating an accelerating universe, the first point to emphazise is that the result (\ref{mu}) is obtained \textit{without having to consider at all such an acceleration of the expansion} (and consequently without having to solve related problems).
Eq.(\ref{c}) in fact leads to the deceleration parameter $q = - \ddot{a}a/\dot{a}^2=0$.\\
It can be also noted that our result does not depend on $\Omega_m$, $\Omega_k$ and $\Omega_{\Lambda}$ and that however it directly gives the expected result without having to adjust any parameter.\\
In the standard model the general expression of the luminosity distance $d_L$ can be expressed in terms of an integral over the redshift $z'$ of the propagating photon as it travels from $z'$ to us at $z'=0$. We thus have in that case:
$$d_L = (1+z)\int_0^z \frac{c\, dz'}{H_0\, E(z')} $$
with
\begin{equation} \label{Pee}
E(z) =\sqrt{\Omega_M (1+z)^3 + \Omega_k (1 + z)^2 + \Omega_{\Lambda}}
\end{equation}
The reason why our eq.(\ref{mu}) does not depend on $\Omega_m$, $\Omega_k$ and $\Omega_{\Lambda}$ comes from the fact that using eq.(\ref{c}) implies a variation with time (more precisely as $a(t)^{-2}$) of the cosmological constant which is not the case for calculations leading to (\ref{Pee}). \\
Of course, our numerical value for $H_0$ is different from the one obtained in the standard cosmology model but it must be underlined that the same observations don't lead, within a given model, to the same numerical values in another one. It would consequently be fallacious to judge a model with the presuppositions of another one.
\section{Conclusion}
At the price of accepting a variation with time of the cosmological "constant" and a density of the universe varying as $a(t)^{-2}$ (so that the mass of the universe varies as $a(t)$), the cosmological model we deduced from eq.(\ref{c}) allows to reduce and enlights a number of problems of the standard cosmology. We know that it may raise questions and so we will present it in more details. \\
However, we cannot think that such an agreement between our eq.(\ref{mu}) and that of Traunmüller \cite{xxxx} and of Marosi \cite{Marosi14} could be a fortuitous coincidence. 
Noting that the reason of such an agreement comes from eq.({\ref{c}}) which gives
\begin{equation}
c \,dt = R(t)\, d\chi \qquad \overset{(\ref{c})}\Longrightarrow \qquad \alpha \,dR(t) = R(t)\, d\chi
\end{equation}
and thus introduces the $ln(\frac{R(t_0)}{R(t_e)})= ln(1 + z)$ in the result, we are forced to consider the interest of eq.(\ref{c}).\\
The constant c was first introduced as the speed of light. However, with the development of 
physics, it came to be understood as playing a more fundamental role, its significance being 
not directly that of a usual velocity (even though its dimensions are) and one might thus think 
of $c$ as being a fundamental constant of the universe. Moreover, the advent of Einsteinian relativity, the fact that $c$ does appear in phenomena where there is neither light nor any motion 
(for example in the fundamental equation $E = m\,c^2$) and its double-interpretation in terms of  velocity of light and of velocity of gravitation forces us to associate it with the 
theoretical description of space-time itself rather than that of the phenomena taking place there
so we have connected the two. For this purpose, we note that there are two universal relations connecting space and time (one is given by the  Einstein's constant $c$ 
and the other appears in the expansion of the universe) and we consider as a logical necessity that there are not two different universal relations between space and time having the same physical dimension of a velocity. We are thus lead to tie the two. Equation (\ref{c}) thus gives a physical interpretation of $c$ and shows that $c$ can be defined solely from the knowledge of the geometry of space-time, that is from its size and its age.\\
Eq.(\ref{c}), thus gives \textit{mathematically} to $c$ the status of a fundamental geometrical constant of the universe, a status that everybody admits without ever giving it any mathematical formulation or any physical origin.
\section{Acknowledgments}
We wish to thank Dr. Harmut Traunmüller for having drawn our attention to the papers of Marosi \cite{Marosi14} and Wei \cite{Wei10}.


\begin{thebibliography}{0}
\bibitem{Vigoureux08} J.M. Vigoureux, P. Vigoureux, B. Vigoureux, Cosmological Applications of a Geometrical Interpretation of "c". \textit{Int. J. Theor Phys.}, \textbf{47}, 928-935, (2008), arXiv:0711.3990 [astro-ph].
\bibitem{Vigoureux03} J.M. Vigoureux, B. Vigoureux, P. Vigoureux, The Einstein constant c in light of Mach's principle, cosmological applications., \textit{Found. Phys. Lett.} 16(2), 183-193 (2003)
\bibitem{Vigoureux} J.M. Vigoureux, B. Vigoureux, M. Langlois, A new Cosmological Model, "Aspects of Today's Cosmology", Antonio Alfonso-Faus ed., ISBN 978-953-307-626-3 
\bibitem{Viennot09} D. Viennot, J.-M. Vigoureux, The Cosmological Constant and the Coincidence Problem in a New Cosmological Interpretation of the Universal Constant “c”, 
\textit{Int. J. Theor. Phys.} (2009) 48: 2246–2252, arXiv:0905.4576 [astro-ph.CO]
\bibitem{xxxx} H. Traunmüller, From Magnitudes and Reshifts of Supernovae, their light curves and angular size of galaxies to a tenable cosmology. Astrophys Space Sci, \textbf{350}, 755-767 (2014).
\bibitem{Marosi14} Laszlo A Marosi ., Hubble diagram Test of 280 Supernovae Redshift Data, \textit{Journal of Modern Physics}, 2014, \textbf{5}, 29-33.
\bibitem{Wei10} H. Wei, Observational constraints on cosmological models with the updated long gamma-ray bursts, \textit{Journal of Cosmology and Astroparticle Physics}, 2010, 1008, 020, arXiv:1004.4951 [astro-ph.CO]
\end{thebibliography}
\end{document}